\documentclass[12pt]{article}

\usepackage{latexsym,amssymb}

\DeclareFontFamily{OT1}{rsfs10}{}
\DeclareFontShape{OT1}{rsfs10}{m}{n}{ <-> rsfs10 }{}
\DeclareMathAlphabet{\mathscript}{OT1}{rsfs10}{m}{n}

\newcommand{\eref}[1]{(\ref{#1})}
\newcommand{\sref}[1]{\ref{#1}}

\newcommand{\cref}[1]{Chapter~\ref{#1}}
\newcommand{\bcenter}{\begin{center}}
\newcommand{\ecenter}{\end{center}}
\newcommand{\beq}{\begin{equation}}
\newcommand{\eeq}{\end{equation}}
\newcommand{\bea}{\begin{eqnarray}}
\newcommand{\eea}{\end{eqnarray}}
\newcommand{\bean}{\begin{eqnarray*}}
\newcommand{\eean}{\end{eqnarray*}}
\newcommand{\ba}{\begin{array}}
\newcommand{\ea}{\end{array}}
\newcommand{\ben}{\begin{enumerate}}
\newcommand{\een}{\end{enumerate}}
\newcommand{\bi}{\begin{itemize}}
\newcommand{\ei}{\end{itemize}}
\newcommand{\bd}{\begin{description}}
\newcommand{\ed}{\end{description}}
\def\fnote#1#2{\begingroup\def\thefootnote{#1}\footnote{#2}
     \addtocounter{footnote}{-1}\endgroup}

\def\End{\mbox{End}}

\def\IZ{\mathbb{Z}}

\def\IP{\mathbb{P}}

\def\dim{{\rm dim}}
\def\cN{{\mathcal N}}
\def\cS{{\mathcal S}}
\def\cO{{\mathcal O}}
\def\cC{{\mathcal C}}
\def\cF{{\mathcal F}}
\def\nn{\nonumber}
\def\td{\mbox{td}}
\def\ch{\mbox{ch}}
\def\rk{\mbox{rk}}

\begin{document}

\begin{titlepage}

\vspace{-2cm}

\title{
   \hfill{\normalsize  UPR-1045T} \\[1em]
   {\LARGE The Moduli of Reducible Vector Bundles}}
\author{Yang-Hui He, Burt A.~Ovrut, and Ren\'{e} Reinbacher
	\fnote{~}{yanghe@physics.upenn.edu, 
	ovrut@physics.upenn.edu,
	rene01@student.physics.upenn.edu}\\[0.5cm]
   {\normalsize Department of Physics, University of Pennsylvania} \\
   {\normalsize Philadelphia, PA 19104--6396, USA}}

\date{}

\maketitle

\begin{abstract}
A procedure for computing the dimensions of the moduli spaces of
reducible, holomorphic vector bundles on elliptically fibered
Calabi-Yau threefolds $X$ is presented. This procedure is applied to
poly-stable rank $n+m$ bundles of the form $V \oplus \pi^* M$, where
$V$ is a stable vector bundle with structure group $SU(n)$ on $X$ and
$M$ is a stable vector bundle with structure group $SU(m)$ on the base
surface $B$ of $X$. Such bundles arise from small instanton
transitions involving five-branes wrapped on fibers of the elliptic
fibration. The structure and physical meaning of these transitions are
discussed.
\end{abstract}

\thispagestyle{empty}

\end{titlepage}


\newpage
\section{Introduction}
When Ho{\v r}ava-Witten theory \cite{HW} is compactified on a smooth
Calabi-Yau threefold one obtains, at intermediate energy scales, a
five-dimensional theory called heterotic M-theory
\cite{Lukas:1998yy,Lukas:1998tt}. Heterotic M-theory is characterized by a
five-dimensional bulk space-time bounded on either side of the
fifth-dimension by four-dimensional ``end-of-the-world'' orbifold
fixed planes. These boundary three-branes descend from Ho{\v r}ava-Witten
theory as the uncompactified part of nine-branes partially wrapped on
the Calabi-Yau threefold.

Before compactification, each nine-brane carries an $\cN = 1$
supersymmetric $E_8$ Yang-Mills supermultiplet on its
worldvolume. Upon compactification, however, a $G$-instanton, that is, a
static solution of the hermitian Yang-Mills equations with structure
group $G \subseteq E_8$, may be present on the Calabi-Yau
threefold. In this case, the associated three-brane carries a reduced
gauge group $H$, which is the commutant of $G$ in $E_8$. In addition,
its worldvolume theory is $\cN = 1$ supersymmetric and contains chiral
fermions transforming in several representations of $H$. Therefore,
the boundary three-branes in heterotic M-theory can potentially carry a
realistic theory of particle physics on their worldvolumes.

To analyze this possibility, one must be able to explicitly construct
these worldvolume theories. It is clear that their structure is
almost
entirely determined by the properties of the $G$-instanton on the
Calabi-Yau threefold. Here, one confronts a seemingly insurmountable
problem, namely that there are no known solutions of the hermitian
Yang-Mills equations on these manifolds. This is due, in part, to the
fact that explicit metrics on Calabi-Yau threefolds also have never
been constructed. However, a way around this problem was found by
Uhlenbeck and Yau \cite{UhYau} and Donaldson \cite{Don}. These authors
showed that stable, holomorphic vector bundles with structure group
$G$ on a Calabi-Yau threefold always admit a unique connection that
must satisfy the hermitian Yang-Mills equations. They also proved the
converse. Therefore, finding a solution of the hermitian Yang-Mills
equations is equivalent to constructing the appropriate vector
bundles. Happily, it is relatively easy to find stable, holomorphic
vector bundles with arbitrary structure group $G$ on Calabi-Yau
threefolds. 

A major step was taken in that direction by Freedman, Morgan,
Witten \cite{FMW1,FMW2,FMW3} and Donagi \cite{Donagi}, who gave a
concrete procedure for constructing such bundles on elliptically
fibered Calabi-Yau threefolds. This work was used, and extended, in a
number of papers
\cite{transition,holo}, where it
was shown that bundles of this type could indeed produce
phenomenologically relevant particle physics theories on a
threefold. Specifically, these papers showed that viable grand unified
theories with gauge groups $E_6, \ SO(10)$ and $SU(5)$, 
for example, can be
so constructed. The results of \cite{transition,holo} 
were limited to Calabi-Yau
threefolds with trivial homotopy. In \cite{SM,SM2,SMnon,SMnon2} and more
recently \cite{Ovrut:2002jk,Ovrut:2003zj,rene3}, 
methods were introduced for
constructing stable, holomorphic vector bundles on torus
fibered Calabi-Yau threefolds with non-trivial homotopy. These results
allow one to add flat bundles, Wilson lines, to the
$G$-instanton. With this extended capability, standard-like models
with gauge group $SU(3)_C \times SU(2)_L \times U(1)_Y$
\cite{SM,SM2,SMnon,SMnon2} and
$SU(3)_C \times SU(2)_L \times U(1)_Y \times U(1)_{B-L}$
\cite{Ovrut:2002jk,Ovrut:2003zj,rene3} have
now been constructed.

There is another interesting generalization of the theory of
holomorphic vector bundles on elliptically
fibered Calabi-Yau threefolds which has been less studied. All of the
above results constructed stable vector bundles with simple structure
groups $G$. However, one can show that so-called poly-stable bundles,
that is, those with semi-simple structure groups such as $G = SU(n)
\times SU(m)$, also admit connections satisfying the hermitian
Yang-Mills equations. These poly-stable bundles were introduced in
this context in \cite{transition}.
Such vector bundles,
having different structure groups, produce a different pattern of
$E_8$ symmetry breaking and, therefore, different low energy gauge
groups $H$. Some breaking patterns in this context were presented in
\cite{transition}. 

Significantly, it was shown in
\cite{transition,inst} that poly-stable holomorphic vector bundles must
always arise in small instanton transitions involving a five-brane
wrapped on a vertical fiber in the elliptically
fibered Calabi-Yau threefold. For example, if prior to a small
instanton transition a three-brane has a stable bundle with structure
group $SU(n)$, then after a transition involving $k$ vertical fibers
the bundle becomes poly-stable with structure group $SU(n) \times
SU(m)$, where $1 < m \le k$. The $SU(m)$ factor of the vector bundle,
the part contributed by the $k$ fibers wrapped by five-branes coalescing
with the initial three-brane, has a very specific structure. It is the
pull-back to the full elliptically
fibered Calabi-Yau threefold of a bundle constructed from $k$ points
on the base surface. Since phase transitions of this
type may arise in a physical context, such as in the Ekpyrotic
theory of Big Bang cosmology \cite{Ek}, poly-stable holomorphic vector
bundles are of potential interest.

In this paper, we will examine a fundamental property of poly-stable
holomorphic vector bundles with structure group $G = SU(n) \times
SU(m)$ on elliptically fibered Calabi-Yau threefolds, namely, the
dimension of their moduli spaces. The structure and dimension of the
moduli spaces of stable vector bundles on these
manifolds, such as the pure $SU(n)$ part of our poly-stable bundle, is
known \cite{moduli}. In this paper, we will extend these results to
compute the dimension of the moduli spaces of stable, holomorphic
bundles with structure group $SU(m)$
on a surface. We will show that this is identical to the
number of moduli of the pull-back
of the bundle constructed from $k$ points on the base of the
Calabi-Yau space. This is the pure $SU(m)$ part of our poly-stable
vector bundle. However, there can be, and there are, moduli associated
with the relationship of the $SU(n)$ and $SU(m)$ bundles to each
other. These moduli are much harder to compute. Be that as it may, we
explicitly perform that calculation in this paper, restricting our
base surface to be $d{\mathbb P}_9$ for concreteness. Putting
everything together, we present a general formula for the dimension of
the moduli space of generic poly-stable holomorphic vector
bundles on elliptically
fibered Calabi-Yau threefolds with base $d{\mathbb P}_9$. This result
is easily generalized to all other base surfaces.

Specifically, in this paper we do the following.
In Section 2, we remind the reader of the structure of holomorphic
vector bundles on elliptically fibered Calabi-Yau threefolds over
$d{\mathbb P}_9$. We also
establish our notation and preparatory formulas which
will be used in the computation. Next, in Section 3, we present the
physical motivation of why we are interested in poly-stable vector
bundles and their relation to small instanton transitions in heterotic
M-theory. Section 4 contains the details of the
calculation of the dimension of the moduli space of these
poly-stable holomorphic vector bundles. Finally, in Section 5, we
present our result, together with a sample calculation. As a check on
the validity of key steps in our computations, alternative proofs in a
restricted context are carried out. These are given in Appendices A
and B. While this paper was in preparation, 
\cite{bjorn} appeared. There is some overlap between the topics
discussed in \cite{bjorn} and the results presented here.

\section{Reducible Bundles on Elliptically Fibered Calabi-Yau Threefolds}
In this section, we briefly review the basic ingredients used in 
our calculation and establish the nomenclature.
In particular, in Sub-section \sref{sec:X},
we concern ourselves with Calabi-Yau threefolds $X$ that are
elliptically fibered over $d{\mathbb P}_9$ surfaces. 
We then review some of the rudiments of the spectral cover
construction of stable holomorphic 
vector bundles $V$ on $X$ in Sub-section \sref{sec:V}. Finally, in
Sub-section \sref{sec:Vhat}, 
we define and discuss the construction of a specific class of 
poly-stable vector bundles $\hat{V}$ on $X$.
\subsection{Elliptically Fibered Calabi-Yau Threefolds $X$ over
$d{\mathbb P}_9$}
\label{sec:X}
We will consider Calabi--Yau threefolds $X$ which are elliptic
curves fibered over a base surface $B$. In other words,
there is a mapping
$\pi: X\to B$ such that $\pi^{-1}(b)$ is a smooth torus, 
$E_b$, for each generic point $b \in B$.
Moreover, this torus fibered threefold
has a zero section. That is, there 
exists an analytic map $\sigma : B\to X$ that assigns to 
every element $b$ of $B$ an element $\sigma(b) \in E_b$. The
point $\sigma(b)$ acts as the zero element for an Abelian group
and turns $E_b$ into an elliptic curve and $X$ into an elliptic
fibration.
Throughout this paper, we will, for concreteness, 
focus on the case where the base $B$ is
a rationally ruled elliptic surface, also called a $d{\mathbb P}_9$
surface. That is,
\beq
B = d\mathbb{P}_9 \ .
\eeq
Such a surface is itself an elliptic
fibration over a base $\IP^1$ with projection map
$\pi_B : d{\mathbb P}_9 \rightarrow \IP^1$ and zero section $\sigma_B :
\IP^1 \rightarrow d{\mathbb P}_9$. Its
structure is that of a
$\IP^2$ blown-up at nine generic points.
In summary, our Calabi-Yau 
threefold $X$ has the structure of a double
elliptic fibration
\beq
X \stackrel{\pi}{\longrightarrow} B 
\stackrel{{\pi_B}}{\longrightarrow}
\IP^1 \ , 
\eeq
with zero sections $\sigma$ and $\sigma_B$
respectively.
We denote the fiber classes of $\pi$ in $X$ and of $\pi_B$ in
$B$ as $F$ and $f$ respectively.

Many properties of $d\mathbb{P}_9$ are described, for
example, in \cite{Ovrut:2002jk}.
Those properties which we will need in this
paper are the following.
First, the second homology group, $H_2(B, \IZ)$, 
is spanned
by $\ell$, the hyperplane class of the $\IP^2$, as well as the nine
exceptional divisors $E_i, {i=1,\ldots, 9}$ of the blow-up.
These classes are effective and have the following intersection numbers
\beq
\label{interdP9}
E_i \cdot E_j = -\delta_{ij}, \qquad E_i \cdot \ell = 0, \qquad
\ell \cdot \ell = 1 \ .
\eeq
Second, the first Chern class of the tangent bundle $TB$ is
\beq
\label{cherndP9}
c_1(TB) = -c_1(K_{B}) = 3 \ell - \sum_{i=1}^9 E_i \ ,
\eeq
where $K_{B}$ is the canonical bundle
\cite{Ovrut:2002jk}. In fact, the first Chern class of the
anti-canonical bundle
$K_B^*$ is precisely the fiber class $f$ of
this fibration over $\IP^1$.
That is,
\beq
\label{dP9K}
-c_1(K_B) = f \ .
\eeq
It then follows from \eref{cherndP9} that
\beq
f = 3 \ell - \sum_{i=1}^9 E_i \ .
\eeq

Recall that a line bundle ${\cal O}_B(D)$ associated with a divisor
$D$ in any surface $B$ is ample, or positive, if 
\beq
\label{ample}
D\cdot D >0, \qquad D \cdot c > 0
\eeq
for every effective irreducible
curve $c$ in $B$.
An effective class in homology simply means that
it is an actual geometric object. For example, an effective class in
$H_2(B,\IZ)$ is a curve. One also called $D$ satisfying \eref{ample}
an ample divisor.
A surface $B$ is called ample if its
anti-canonical bundle $K_B^*$ is. 
We will often
refer to these conditions of ampleness and effectiveness.
Now note, using \eref{interdP9} and \eref{cherndP9}, 
that
\beq
c_1(K_B)^2 = 0 \ .
\eeq
Therefore, it follows from \eref{ample} that
$K_B^*$ and, hence, $d{\mathbb P}_9$, is not ample. This is
why a $d{\mathbb P}_9$ surface is sometimes called a half-K3. See, for
example, \cite{review}.
Next, let us consider $c_2(TB)$, the highest 
Chern class on $d{\mathbb P}_9$. This is nothing but the
topological
Euler characteristic, $\chi(d{\mathbb P}_9) = 12$. See, for
example, \cite{Ovrut:2002jk}. Therefore,
\beq
\label{c2dP9}
c_2(TB) = 12 \ .
\eeq
We will also need the following intersection numbers,
\beq
\label{intersect}
\sigma_B \cdot \sigma_B = -1, \qquad \sigma_B \cdot f = 1, 
\qquad f \cdot f = 0 \ .
\eeq
The first intersection follows simply from the adjunction formula and
\eref{dP9K}.
The second equality follows because we are
intersecting the fiber with a section, and the last because we are
intersecting a fiber with itself. 

Third, we need to know 
the effective curves in $B$. The irreducible
effective curves in $d{\mathbb P}_9$ are spanned by \cite{Ron}
$f = 3 \ell - \sum_{i=1}^9 E_i$ and the curves 
\beq
\label{fy}
y = \beta \ell + \sum_{i=1}^9 \alpha_i E_i
\eeq
with
\beq
\alpha_i \in \IZ, \qquad \beta \in \IZ_{\ge 0} \ ,
\eeq
satisfying the constraints
\beq
y^2 = -1, \qquad y \cdot f = 1 \ .
\eeq
These constraints imply that
\beq
\label{alphabeta}
-\beta^2 + \sum\limits_{i=1}^9 \alpha_i^2 = 1, \qquad 
3 \beta + \sum\limits_{i=1}^9 \alpha_i = 1 \ .
\eeq

Finally, 
the Chern classes for the elliptically fibered Calabi-Yau threefold 
$X$ were presented in \cite{holo}. We will
make use of them later. Note that 
\beq
c_1(TX) = 0 \ ,
\eeq
since $X$ is a Calabi-Yau manifold. The second Chern class is given by
\beq
c_2(TX) = 12 \sigma
\cdot \pi^*(c_1(TB)) + \pi^*(c_2(TB) + 11 c_1^2(TB)) \ . 
\eeq
In our case, $B = d{\mathbb P}_9$. Then,
it follows from \eref{cherndP9}, \eref{dP9K} and \eref{c2dP9} that 
\beq
\label{chernX}
c_2(TX) = 12 \sigma \cdot \pi^*(f) + 12 F \ .
\eeq
\subsection{Spectral Cover Construction of Stable $SU(n)$ Bundles over $X$}
\label{sec:V}
One ingredient we need in this paper is stable,
holomorphic $SU(n)$ vector bundles over the
Calabi--Yau threefold
$X$. We will construct such bundles using the so-called
spectral cover method \cite{FMW1,FMW2,FMW3,Donagi}. We adhere to
the notation and presentation in \cite{transition}.
Stable, holomorphic $SU(n)$ vector 
bundles $V$ over $X$ can be explicitly constructed from two
objects, a divisor $\cC_X$ of $X$, called the spectral cover, and a
line bundle $\cN_X$ on $\cC_X$. The pair $(\cC_X,\cN_X)$ is called the 
spectral data for $V$.

The spectral cover, $\cC_X$, is a surface in
$X$ that is an $n$-fold cover of the base $B$. Its general form is
\beq
\label{coverX}
\cC_X=n\sigma + \pi^*\eta \ ,
\eeq
where $\sigma$ is the zero section 
and $\eta$ is a curve in $B$. 
If $\cC_X$ is an effective, irreducible surface, then the associated
vector bundle is stable. We will always impose this restriction.
In addition to $\cC_X$, we must also specify a line bundle $\cN_X$. 
For $SU(n)$ vector bundles, this must satisfy
\begin{equation}
c_{1}(\cN_X)=n(\frac{1}{2}+\lambda)\sigma+(\frac{1}{2}-\lambda)
\pi^{*}\eta+(\frac{1}{2}+n\lambda)\pi^{*}c_{1}(TB) \ ,
\end{equation}
where $\lambda$ is a rational number such that
\beq
\label{lambda}
\ba{rcll}
\lambda &=& p + \frac12, \qquad &n \mbox{~~odd} \\
\lambda &=& p, \qquad & n \mbox{~~even}
\ea
\eeq
and $p \in \IZ$.

Finally, the stable $SU(n)$ vector bundle $V$ on $X$ is constructed
from the spectral data $(\cC_X,\cN_X)$ as
$V=\pi_{1*}(\pi^{*}_{2}\cN_X \otimes {\mathcal P})$,
where $\pi_{1}$ and $\pi_{2}$ are the projections of the fiber
product $X \times_{B} \cC_X$ onto the two factors $X$ and $\cC_X$
respectively and
${\mathcal P}$ is the associated Poincar\'{e} bundle. We
refer the  reader to \cite{FMW1,holo} for a detailed discussion.
This is essentially the Fourier-Mukai transform, 
which works in reverse as
well. In other words, there is a 1-1 correspondence between $V$ and the
spectral data
\beq
\label{FMV}
(\cC_X,\cN_X) \stackrel{\mbox{Fourier-Mukai}}{\longleftrightarrow} V \ .
\eeq
For $SU(n)$ bundles, 
\beq
c_1(V)=0 \ .
\eeq
The remaining
Chern classes were computed in \cite{FMW1,holo} and found to be
\beq
c_2(V)=\sigma \cdot \pi^*(\eta) -
	\frac{1}{24}c_1(TB)^2(n^3-n)
	+\frac{1}{2}(\lambda^2-\frac{1}{4})n \ 
	\eta \cdot (\eta-nc_1(TB)) F, \quad
c_3(V)= 2\lambda \eta \cdot (\eta-n c_1(TB)) \ .
\eeq
For $B=d{\mathbb P}_9$, 
these simplify by virtue of \eref{dP9K} and \eref{intersect}. We find
that
\beq
\label{chernV}
c_2(V)=\sigma \cdot \pi^*(\eta) +
	\frac{1}{2}(\lambda^2-\frac{1}{4})n\eta \cdot (\eta-nf) F ,
	\quad 
c_3(V)= 2\lambda \eta \cdot (\eta- n f) \ .
\eeq

Let us consider the structure of the spectral cover $\cC_X$ for a 
$d{\mathbb P}_9$ base surface. As stated previously, $\cC_X$ must be
both effective and irreducible. These requirements put constraints on
the curve $\eta$ which, in the $d{\mathbb P}_9$ context, are
complicated to solve. Happily, we need not do this in general. As we
will see in Section 4, it is expedient to restrict $\cC_X$ to be an
ample divisor in $X$. $\cC_X$ being ample immediately implies that it
is effective and simplifies the proof that it is irreducible. As we
will now show, the conditions for $\cC_X$ to be ample and irreducible
are relatively straightforward.

If $\cC_X$ is ample, then so is $\eta$. Unfortunately, the converse is
not true. However, we can prove a partial converse. Let $\eta'$ be any
ample curve in $d{\mathbb P}_9$ and consider
\beq
\label{etaconv}
\eta = \eta' + (n+1) f \ .
\eeq
Clearly, $\eta$ is also an ample curve. Note, however, that not every
ample curve in $d{\mathbb P}_9$ is of the form \eref{etaconv}. We can
show that for every ample curve of the form \eref{etaconv}, $\cC_X$ is
both ample and irreducible, as required. For $\eta'$ to be ample in 
$d{\mathbb P}_9$, it must satisfy
\beq
\label{etaample1}
\eta' \cdot \eta' > 0
\eeq
and its intersection with all effective curves must be positive. That
is,
\beq
\label{etaample2}
\eta' \cdot f > 0, \quad
\eta' \cdot y > 0 \ .
\eeq
Expanding $\eta'$ in the basis of $H_2(B, \IZ)$ as
\beq
\eta' = b' \ell + \sum\limits_{i=1}^9 a_i' E_i \ , 
	\qquad a_i', b' \in \IZ \ , 
\eeq
and using
\beq
\label{etadotprime}
\eta' \cdot \eta' = b'^2 -\sum_{i=1}^9 a_i'^2 \ , \qquad 
\eta' \cdot f = 3b' + \sum_{i=1}^9 a_i' \ ,
\eeq
we see that \eref{etaample1} and \eref{etaample2} 
translate into three
constraints on the coefficients $a_i'$ and $b'$. These are
\beq
\label{positive}
b'^2-\sum_{i=1}^9 a_i'^2 > 0, \qquad
3b'+ \sum\limits_{i=1}^9 a_i' > 0, \qquad
b'\beta - \sum\limits_{i=1}^9 a_i' \alpha_i > 0
\eeq
for all $\alpha_i \in \IZ$ and $\beta \in \IZ_{\ge 0}$ satisfying
\eref{alphabeta}.
If we also expand $\eta$ into the basis of $H_2(B, \IZ)$ as
\beq
\label{etaexp}
\eta = b \ell + \sum\limits_{i=1}^9 a_i E_i \ , \qquad a_i, b \in
\IZ \ , 
\eeq
then it follows from \eref{etaconv} that
\beq
\label{convprime}
a_i = a_i' - (n+1), \quad b = b' + 3(n+1) \ .
\eeq

To conclude this section, 
we present some useful intersection formulas which we will use
frequently throughout this paper \cite{transition}. First, we have
\beq
\label{etadot}
\eta \cdot \eta = -\sum_{i=1}^9 a_i^2 + b^2, \qquad 
\eta \cdot c_1(TB) = \eta \cdot f = 3b + \sum_{i=1}^9 a_i \ .
\eeq
Next, we have
\beq
\label{sigsig1}
\sigma \cdot \sigma = -(\pi^*(c_1(TB))) \cdot \sigma = - \pi^*(f)
\cdot \sigma \ .
\eeq
Furthermore, note, using \eref{cherndP9}, that the triple intersection,
\bea
\sigma^3 &=& (- \pi^*(c_1(TB)) \cdot \sigma ) \cdot \sigma
= (- \pi^*(c_1(TB)) \cdot (- \pi^*(c_1(TB))) \cdot \sigma \nonumber \\
&=& \pi^*(c_1(TB) \cdot c_1(TB)) \cdot \sigma 
= 0  \ .
\label{sig3}
\eea
Finally, since we can always choose a
representative $F$ of the fiber of $X$ which will not intersect the
pull-back of the curve $\eta$, we have 
\beq
\label{Feta}
F \cdot \pi^*(\eta) = 0 \ .
\eeq
Also, it is obvious, since $\sigma$ is a section, that
\beq
\label{Fsigma}
F \cdot \sigma = 1 \ .
\eeq

\subsection{Reducible Vector Bundles on $X$}
\label{sec:Vhat}
The previous sub-section reviewed the spectral cover 
construction of stable, holomorphic 
$SU(n)$ vector bundles $V$ over elliptically
fibered Calabi-Yau
threefolds $X$ with base $B=d{\mathbb P}_9$. 
Our main concern in this paper will be
reducible, poly-stable rank $n+m$ holomorphic
vector bundles over the same space $X$. In other
words, we want to construct bundles with structure group
$SU(n) \times SU(m)$. For physical reasons to be discussed in the next
section, we will restrict the bundles associated with the factor group
$SU(m)$ to be the pull-back to $X$ of stable rank $m$ holomorphic
vector bundles on the $d{\mathbb P}_9$ base surface.
This
construction was carried out in Section 6 of \cite{transition}.

\subsubsection{$d{\mathbb P}_9$ Bundles and Their Pull-back to $X$}
We want to construct
stable, holomorphic $SU(m)$ vector bundles $M$ with $m\geq2$
over the base $B$ with the Chern classes given by
\beq
\label{chernM}
c_{1}(M)=0, \quad c_{2}(M)=k \in {\mathbb Z}, 
\quad c_3(M)=0
\ .
\eeq
In this paper,
the base is $d{\mathbb P}_9$ 
with a zero section $\sigma_B$. 
Hence, we can use the spectral cover construction 
with spectral data $({\cal C}_B, {\cal N}_B)$. Here, 
${\cal C}_B$ is a
curve in $B$ which is an $m$-fold cover of the base $\IP^1$.
Generically, it is given by
\beq
\label{coverbase}
{\cal C}_B = m \sigma_B + k f \ .
\eeq
Recall that $\cC_B$ must be both effective and irreducible in order
for $M$ to be stable. $\cC_B$ will be effective if we choose
\beq
k \ge 0 \ .
\eeq
To ensure that the spectral cover is irreducible, 
we must impose \cite{transition} the constraint that
\beq
\label{km}
1 < m \le k \ .
\eeq
${\cal N}_B$ is the spectral
line bundle over ${\cal C}_B$. In order for the vector bundle $M$
corresponding to $({\cal C}_B, {\cal N}_B)$ to have Chern classes
\eref{chernM}, we need to require that
\cite{transition}
\beq
\label{NBase}
c_1(\cN_B) = \frac12 m (2k -1 - m) \ .
\eeq
One can now use \eref{coverbase} and \eref{NBase} to construct the
requisite vector bundles via the Fourier-Mukai transformation
Schematically, we have the structure
\beq
\label{fiberdP9}
({\cal C}_B, {\cal N}_B)
\stackrel{\mbox{Fourier-Mukai}}{\longleftrightarrow} M \ .
\eeq

Having constructed the bundles $M$ on $d{\mathbb P}_9$, we can easily
lift them to the Calabi--Yau threefold $X$.
The pull--back $\pi^*M$ of $M$ to $X$ is a stable holomorphic
$SU(m)$ vector bundle with $m\ge 2$ over $X$.
Its Chern classes are
\beq
c_{1}(\pi^{*}M)=0, \quad c_{2}(\pi^{*}M)=kF, 
\quad c_3(\pi^{*}M)=0 \ . 
\eeq

\subsubsection{Reducible $SU(n)\times SU(m)$ Bundles over $X$}
Equipped with two stable holomorphic
vector bundles over $X$, namely the $SU(n)$
bundle $V$ in \eref{FMV} and the $SU(m)$ bundle $\pi^*M$, which is
a pull-back of the bundle $M$ in \eref{fiberdP9} on
$d{\mathbb P}_9$, our
preparatory work is done. We can now construct reducible, poly--stable, 
rank $n+m$ holomorphic
vector bundles with structure group $SU(n)\times
SU(m)$ over $X$, which we denote by $\hat{V}$, simply as the direct sum
\beq
\hat{V} = V \oplus \pi^*M \ .
\eeq
The Chern classes of this reducible bundle are easily computed to be
\beq
c_1(\hat{V}) = 0, \quad 
c_2(\hat{V}) = c_2(V) + k F, \quad
c_3(\hat{V}) = c_3(V) \ ,
\eeq
where $c_2(V)$ and $c_3(V)$ were given in \eref{chernV}.
Schematically, the relation between the Calabi-Yau 
threefold $X$, the base $B =
d{\mathbb P}_9$, and the bundles is
\beq
\ba{cclc}
& V \oplus \pi^* M & & M \\
& \downarrow & & \downarrow \\
\pi: & X & \longrightarrow & B
\ea
\label{fiberCY}
\eeq

\section{Small Instanton Transitions}
Having constructed 
reducible $SU(n) \times SU(m)$ vector bundles on $X$, we now
recall why such objects are of interest to us. 
The Ho\v rava-Witten vacuum \cite{HW}
of M-theory is an $S^1/\IZ_2$ interval with an 11-dimensional bulk
space and
two ``end-of-the-world'' 10-dimensional fixed planes, each carrying an
$\cN =1$ $E_8$ Yang-Mills supermultiplet. This theory has
chiral fermions on the orbifold planes. 
This vacuum can be further compactified on
a Calabi-Yau threefold, leading to a five-dimensional
``brane-world'' scenario
wherein one fixed plane, or three-brane,
is our ``observable'' 4-dimensional world and
the other is a ``hidden'' brane. This compactified 
theory is called heterotic M-theory \cite{transition}.

A wrapped BPS five--brane in the bulk space of heterotic M-theory
has a modulus corresponding to the 
translation of the five--brane in the orbifold direction. The
following question was addressed in \cite{transition}.
What happens to a wrapped bulk five--brane in heterotic
M--theory when it is
translated across the bulk space and comes into direct contact with
one of the boundary three--branes?
It was shown that in collisions of a bulk five--brane with
the observable boundary three--brane, the 
wrapped five--brane disappears and its data is ``absorbed'' into a
singular bundle, called a torsion free sheaf. This sheaf is 
localized on the Calabi--Yau
threefold associated with the observable three--brane and
is referred to as a ``small
instanton''\cite{inst}. This small instanton can
then be ``smoothed out'' to a
non-singular holomorphic vector bundle by moving in its moduli space.  
The physical picture is that the bulk five--brane
disappears after the collision, but at the cost of 
altering the topology of the 
instanton vacuum on the boundary three--brane. 

There are two different phase transitions associated with the
above-mentioned collisions. First, there is
a ``chirality-changing'' transition, where the number of
lepton-quark families changes and second,
there is a ``gauge-changing'' transition, where the
gauge group on the boundary three-brane is altered. 
Let us briefly remind ourselves of the two transitions.
Recall from
\cite{transition} that for
an elliptically fibered Calabi-Yau threefold 
with zero section $\sigma$ and
fiber class $F$, the bulk five--brane wraps a class
\beq
W = W_B \sigma + a_F F \ ,
\eeq
where $W_B$ is the lift of a curve in the base and the fiber
coefficient $a_F$ is a non-negative integer.
The chirality-changing transitions have the property that they 
absorb all, or part, of the base
component, $W_{B}$, into the holomorphic vector bundle.
Such transitions do not affect the fiber component, $a_{F}F$, of the
five--brane curve, which is identical on either side of the
small instanton transition. On the other hand, in gauge-changing
transitions, all, or a portion, of $a_{F}F$ 
is absorbed via the small instanton
phase transition into the vector bundle on the observable brane.
In these transitions, it is the base component $W_B$ of the
five-brane curve that is left undisturbed. It was shown in
\cite{transition} that these two processes lead to a different kind of
holomorphic vector bundle after the small instanton transition. Let us
assume that initially, prior to the collision,
the observable brane has a stable holomorphic $SU(n)$ vector bundle
$V$ associated with the spectral data $(\cC_X,\cN_X)$. After the
transition, the vector bundle becomes
$V'$ with spectral data
$(\cC'_X,\cN'_X)$. As demonstrated in \cite{transition}, in
chirality-changing transitions the holomorphic vector bundle $V'$ is
stable with structure group $SU(n)$, exactly as for $V$. However, its
spectral cover is altered by absorbing $W_B$ in such a way as to
induce a change in the third Chern class and, hence, in the number of
lepton-quark families. On the other hand, gauge-changing transitions
preserve the third Chern class. However, absorbing $kF$, where $k \le
a_F$, makes $V'$ a reducible bundle with structure group $SU(n) \times
SU(m)$. Furthermore, the bundles associated with the factor group
$SU(m)$ are always a pull-back to $X$ of stable rank $m$ holomorphic
vector bundles on the base. These bundles are constructed from $k$
points, the projection onto the base $B$ of the fibers $kF$, and
satisfy $1 < m \le k$. Both kinds of phase transitions alter the
second Chern class of the holomorphic bundle in such a way as to
preserve the over-all anomaly freedom. The structure of the spectral
data, both before and after a small instanton transition, is listed in
Table \ref{tab:trans}.
\begin{table}[ht]
\label{tab:trans}
\[
\ba{|c|c|}
\hline
\mbox{Before} & \mbox{After} \\
\hline
SU(n), V \leftrightarrow (\cC_X,\cN_X) &
\ba{c|c}
\mbox{Chirality Changing} & SU(n), V' \leftrightarrow 
	(\cC_X \cup W_B, \cN_X) \\
\hline
\mbox{Gauge Changing} & SU(n) \times SU(m), V' \leftrightarrow 
	(\cC_X \cup m \sigma, \cN_X \oplus \sigma_* M) \\
\ea
\\
\hline
\ea
\]
\caption{Vector bundles and the associated spectral data before and after
the collision of the bulk five-brane.}
\end{table}

In this paper, we will focus on gauge changing phase transitions, that
is, the absorption via a small instanton phase transition
of all, or a portion, of $a_F F$ by
the vector bundle on the observable brane. This creates an $SU(n)
\times SU(m)$ vector bundle that is reducible and poly-stable.
This $SU(n) \times SU(m)$ bundle is  
precisely $\hat{V} = V \oplus \pi^* M$ 
introduced in \cite{transition} and reviewed earlier in
\eref{fiberCY} . The above discussions explains the physical relevance
of these poly-stable bundles.
We now proceed to compute
their moduli.
\section{Computing the Moduli of the Reducible Bundle}
In this section, we compute the moduli for any reducible
vector bundle $\hat{V} = V \oplus \pi^* M$, specifically,
the dimension its space of deformations. The space of
deformations of an arbitrary 
vector bundle $U$ on a complex manifold $X$ is given
by \cite{griffith}
\beq
H^1(X, \End(U)) \ ,
\eeq
where 
\beq
\End(U) = U \otimes U^*
\eeq
is the sheaf of 
endomorphisms of $U$.
Therefore, in this paper, we wish to calculate 
\beq
\label{mod}
h^1(X, \End \hat{V}) \ ,
\eeq
with $\hat{V} = V \oplus \pi^*M$.
We can readily express $H^1(X, \End(V))$ as four terms
\beq
\label{H1}
H^1(X, \End \hat{V})
= H^1(X, \hat{V} \otimes \hat{V}^*)
= I\oplus II \oplus III \oplus IV \ .
\eeq
Terms I and IV are
the moduli spaces for the bundles $V$ and $\pi^*M$ respectively, and
are defined as
\beq
I = H^1(X, V \otimes V^*), \quad 
IV = H^1(X,\pi^*M \otimes (\pi^*M)^*) \ .
\eeq
On the other hand, terms II and III are given by
\beq
\label{defII-III}
II = H^1(X, V \otimes (\pi^*M)^*), \quad
III = H^1(X, \pi^*M \otimes V^*)
\eeq
and 
contain the moduli associated with the cross terms between $V$ and
$\pi^*M$. 
We proceed, then, to calculate the dimensions of the
four terms I, II, III and IV.
%
\subsection{Term I: Moduli From $V$}
We are familiar with the first term $I = H^1(X, V \otimes V^*)$.
It contains the
moduli associated with the stable holomorphic $SU(n)$ vector bundle
$V$.
This case was addressed in Section 4 of \cite{moduli}. 
It was shown that with respect to the spectral data $(\cC_X,\cN_X)$,
\beq
\label{dimI}
\dim (I) = h^0(X, {\cal O}_X({\cal C}_X)) - 1 \ .
\eeq
Let us assume that ${\cal O}_X({\cal C}_X)$ 
is positive, as in \cite{moduli}. This condition means that ${\cal
C}_X$ is ample which, as discussed earlier, can be implemented by
imposing the positivity
constraints \eref{positive} and \eref{convprime}.
We can now
easily evaluate \eref{dimI} using the methods introduced in Section 4
of \cite{moduli}.  With the positivity assumption,
all of the higher cohomology classes are zero by
the Kodaira-Serre vanishing theorem and we have
\bea
h^0(X, {\cal O}_X({\cal C}_X)) &=& \chi(X, {\cal O}_X({\cal C}_X))
\nonumber \\
&=& \int_X \mbox{ch}({\cal O}_X({\cal C}_X)) \wedge \td(TX) \nn \\
&=& \frac16 \int_X c_1^3({\cal O}_X({\cal C}_X)) + \frac{1}{12} \int_X
c_1({\cal O}_X({\cal C}_X)) \wedge c_2(TX) \nonumber \\
&=& \frac16 \int_X {\cal C}_X^3 + \frac{1}{12} \int_X {\cal C}_X \wedge 
\left(12 \sigma \cdot \pi^*(f) + 12 F
\right) \label{h0I} \ .
\eea
In evaluating this expression,
we have used the fact that $c_1({\cal O}_X({\cal C}_X)) = {\cal
C}_X$,  the Atiyah-Singer index theorem
and the result for $c_2(TX)$ given in \eref{chernX}.

Finally, recalling that the pull-back of a point $b \in B$ gives the
fiber class $F$, that is, $\pi^*(b) = F$ and using \eref{etaexp},
\eref{sig3}, \eref{Feta} and
\eref{Fsigma},
we can immediately finish the computation of \eref{h0I}.
We find that
\bea
h^0(X, {\cal O}_X({\cal C}_X))
&=&
(1 - \frac12 n^2) \sum_{i=1}^9 a_i
- \frac12 n \sum_{i=1}^9 a_i^2 +
(3 -\frac32 n^2) b
+ \frac12 n b^2
+ n \ .
\eea
It follows that
\beq
\label{termI}
\dim(I) = (1 - \frac12 n^2) \sum_{i=1}^9 a_i
- \frac12 n \sum_{i=1}^9 a_i^2 +
(3 -\frac32 n^2) b
+ \frac12 n b^2 + n -1 \ ,
\eeq
where $a_i$ and $b$ satisfy the conditions given in 
\eref{positive} and \eref{convprime}.

\subsection{Term IV: Moduli from $\pi^*M$}
Next, we consider the term $IV = H^1(X,\pi^*M \otimes
(\pi^*M)^*)$. This corresponds to
the moduli associated with the stable $SU(m)$ vector
bundle $M$ on the base pulled back to $X$.
We will show that this term essentially reduces to the deformations
of $M$ over the base $B$. 

We evoke a Leray spectral sequence in this
context \cite{griffith}. This states that for $\pi: X
\rightarrow B$ and any
vector bundle $U$ on $X$, we have the exact sequence
\beq
0 \rightarrow H^1(B, \pi_* (U \otimes U^*)) \rightarrow H^1(X, U
\otimes U^*)
\rightarrow H^0(B, R^1\pi_* (U \otimes U^*)) \rightarrow  \ldots 
\eeq
where $R^i \pi_*$ is the $i$-th right derived functor for the
push-forward map $\pi_*$. See, for example, \cite{hart1}.
In our case, taking $U$ to be $\pi^*M$, we have the sequence 
\beq
\label{1seqIV}
0 \rightarrow H^1(B, \pi_* (\pi^* M \otimes (\pi^*M)^*)
\rightarrow 
IV
\rightarrow 
H^0(B, R^1\pi_* (\pi^* M \otimes (\pi^*M)^*)) \rightarrow  \ldots 
\eeq
We first use the projection formula \cite{hart1} to simplify this
expression.  This formula states that for a morphism
$f : X \rightarrow Y$, $F$ an ${\cal O}_X$-module and $E$ a 
locally free ${\cal O}_Y$-module of finite rank, we have
\beq
R^if_*(F \otimes f^* E)
\simeq R^if_*(F) \otimes E \ . 
\eeq
Using this and the fact that $(\pi^*M)^* = \pi^*M^*$,
we can write 
\bea
R^1 \pi_* (\pi^*M \otimes (\pi^*M)^*) &=&
	R^1 \pi_* (\pi^*M) \otimes M^* \nonumber \\
&=& R^1 \pi_* ({\cal O}_X \otimes \pi^*M) \otimes M^* \ ,
\eea
where we have tensored with the trivial sheaf ${\cal O}_X$.
We can use the projection formula again to further
reduce this to
\beq
R^1 \pi_* (\pi^*M \otimes (\pi^*M)^*) =
R^1 \pi_* {\cal O}_X \otimes M \otimes M^* \ .
\eeq
Similarly,
\beq
\pi_* (\pi^*M \otimes (\pi^*M)^*) = M \otimes M^*
\eeq
and the spectral sequence \eref{1seqIV} becomes
\beq
\label{seqIV}
0 \rightarrow H^1(B, M \otimes M^*)
\rightarrow 
IV
\rightarrow 
H^0(B, R^1 \pi_* {\cal O}_X \otimes M \otimes M^*) \rightarrow
\ldots
\eeq

We proceed to compute term IV by first focusing on the third term 
in \eref{seqIV} and showing that it is, in fact, zero.
To do this, we evoke relative duality 
\cite{hart2}. This states that for $\pi: X
\rightarrow B$ and any sheaf $\cS$ on $X$, we have
\beq
(R^1 \pi_* \cS)^* \simeq R^0 \pi_*(\cS \otimes K_X \otimes
\pi^* K_B^*) \ .
\eeq
Taking $\cS$ to be ${\cal O}_X$, and recalling that $K_X$ is trivial
since $X$ is a Calabi-Yau manifold, we find
\bea
(R^1 \pi_* {\cal O}_X)^* &=& R^0\pi_* ({\cal O}_X \otimes
	{\cal O}_X \otimes\pi^* K_B^*) \nonumber \\
&=& \pi_* (\pi^* K_B^*) \nn \\
&=& K_B^* \ .
\eea
Therefore, the third term in \eref{seqIV} becomes
\beq
\label{thirdIV}
H^0(B, K_B \otimes M \otimes M^*) \ ,
\eeq
which are the global sections of the bundle $K_B \otimes M \otimes
M^*$. They correspond to global holomorphic maps from
${\cal O}_B \rightarrow K_B \otimes M \otimes M^*$, 
or, equivalently, to maps
\beq
K_B^* \rightarrow M \otimes M^* \ .
\label{kbmap}
\eeq
Now $M$ and, hence, $M^*$ are stable bundles of slope zero. 
Thus, their tensor product is
poly-stable \cite{HL}.
Now for the surface $d{\mathbb P}_9$, $K_B^*$ is
effective\footnote{For all other del Pezzo surfaces, $K_B^*$ is ample
	and a similar argument holds \cite{review}.}. 
If there were any non-trivial maps \eref{kbmap}
we would have a positive sub-bundle of $M \otimes M^*$, thus
violating poly-stability.
Therefore, we conclude that \eref{thirdIV}, that is,
the third term in the sequence \eref{seqIV}, vanishes. 
It follows that all
moduli contributions to IV arise only from the base.
Indeed, the sequence \eref{seqIV} now gives us the isomorphism
\beq
IV \simeq  H^1(B, M \otimes M^*) \ .
\eeq
We have, therefore, reduced the computation of term IV to that of the
moduli 
of a stable rank $m$ vector bundle over the base $B$. 

The computation proceeds similarly to the one in Section 4 of
\cite{moduli}. Recalling the structure of the bundle $M$ from
\eref{fiberdP9}, it is clear that the moduli of $M$ arise
from both the spectral
curve ${\cal C}_B$ and the spectral line bundle ${\cal N}_B$.
The number of such moduli was shown to be
\beq
\label{Bmoduli}
\dim(IV) = [h^0(B, {\cal O}_B({\cal C}_B)) - 1] + h^1({\cal C}_B, {\cal
	O}_{{\cal C}_B}) 
\eeq
in \cite{moduli}, where the term in the square brackets describes
moduli coming from 
${\cal C}_B$ while the last term describes those coming 
from ${\cal N}_B$.
The first term is in analogy with \eref{dimI}. Here, however, the
spectral line bundle $\cN_B$ also contributes moduli.
Thus, we need to compute the
two terms, $h^0(B,{\cal O}_B({\cal C}_B))$ and
$h^1({\cal C}_B, {\cal O}_{{\cal C}_B})$.
\subsubsection{Spectral Curve Moduli}
Let us first compute $h^0(B,{\cal O}_B({\cal C}_B))$, the number of 
spectral
curve moduli. Although we found it expedient to assume that
${\cal O}_X(\cC_X)$ was ample in our computation of term I, we need
not make this assumption for $\cO_B(\cC_B)$. However, we must require 
that $\cC_B$ be irreducible which, by Bertini's Theorem,
requires the constraint \eref{km}.
For general 
${\cal O}_B({\cal C}_B)$, the
following calculation is somewhat technical.
However, the special case where
${\cal O}_B({\cal C}_B)$ is ample can be readily computed along the
same lines as in \cite{moduli}. We present the computation for this 
simple and illustrative case in Appendix A. At the end of this
sub-section, when we find the result for the general case, we will
compare it
with the expression given in Appendix A and see that they agree.
Now
\beq
\label{h0push}
H^0(B,{\cal O}_B({\cal C}_B)) = H^0(\IP^1, \pi_{B*} {\cal
O}_B({\cal C}_B)) \ ,
\eeq
where we have pushed down to the base $\IP^1$.
Recalling from \eref{coverbase} that ${\cal C}_B = m\sigma_B + k
f$, it follows that
\bea
{\cal O}_B({\cal C}_B) &=& {\cal O}_B(m\sigma_B) \otimes 
{\cal O}_B(k f) \nn \\
&=& {\cal O}_B(m\sigma_B) \otimes \pi^* {\cal O}_{\IP^1}(k) \ .
\eea
Using the projection formula again, we obtain
\beq
\label{piBCB}
\pi_{B*} {\cal O}_B({\cal C}_B) = \pi_{B*} {\cal O}_B(m\sigma_B) \otimes
{\cal O}_{\IP^1}(k) \ .
\eeq
Let us focus on the first term $\pi_{B*} {\cal O}_B(m\sigma_B)$. 
The second term is the familar line bundle on $\IP^1$.
We can proceed inductively. First, we note that we have
the exact sequence
\beq
0 \rightarrow {\cal O}_B \rightarrow {\cal O}_B(\sigma) 
\rightarrow {\cal O}_{\sigma}(\sigma) \rightarrow 0 \ .
\eeq
Applying the functor $\pi_{B*}$, we find
\beq
\label{nsig1}
0 \rightarrow \pi_{B*} {\cal O}_B \rightarrow 
\pi_{B*} {\cal O}_B(\sigma)  
\rightarrow \pi_{B*} {\cal O}_{\sigma}(\sigma) \rightarrow 
R^1\pi_{B*} {\cal O}_B \rightarrow R^1\pi_{B*} {\cal O}_B(\sigma)
\rightarrow
\ldots 
\eeq
This will be the starting point of a calculation by induction.
We note that
\beq
\pi_{B*} {\cal O}_B = {\cal O}_{\IP^1}
\eeq
on the base $\IP^1$. Also, we have
\beq
R^1\pi_{B*} {\cal O}_B(\sigma) = 0 \ .  
\eeq
This follows from the Kodaira vanishing theorem 
since ${\cal O}_B(\sigma)$ is of
positive degree. In fact, we have
\beq
\label{r1-1}
R^1\pi_{B*} {\cal O}_B(n \sigma) = 0, \quad n \in \IZ_{>0} \ .
\eeq
Furthermore, it follows from relative duality \cite{hart2} that
\beq
\label{r1-2}
R^1\pi_{B*} {\cal O}_B = {\cal O}_{\IP^1}(-1) \ . 
\eeq
Also, note that
\beq
\label{pi-1}
\pi_{B*} {\cal O}_{\sigma}(\sigma) = {\cal O}_{\IP^1}(-1) \ .
\eeq
In general, we find that
\beq
\label{pi-2}
\pi_{B*} {\cal O}_\sigma(n \sigma) = {\cal O}_{\IP^1}(-n), \quad n \in
\IZ_{>0} \ .
\eeq
Using \eref{r1-1}, \eref{r1-2} and \eref{pi-1},
the sequence \eref{nsig1} becomes
\beq
0 \rightarrow {\cal O}_{\IP^1} \rightarrow
\pi_{B*} {\cal O}_B(\sigma)  
\rightarrow {\cal O}_{\IP^1}(-1) \rightarrow 
{\cal O}_{\IP^1}(-1) \rightarrow 0 \ ,
\eeq
which implies that
\beq
\label{induct1}
\pi_{B*} {\cal O}_B(\sigma) = {\cal O}_{\IP^1} \ .
\eeq
We now proceed to the next short exact sequence in our induction. It
is
\beq
0 \rightarrow {\cal O}_B(\sigma) \rightarrow {\cal O}_B(2\sigma) 
\rightarrow {\cal O}_{\sigma}(2\sigma) \rightarrow 0 \ ,
\eeq
from which we have the long exact sequence
\beq
0 \rightarrow \pi_{B*} {\cal O}_B(\sigma) \rightarrow 
\pi_{B*} {\cal O}_B(2 \sigma)  \rightarrow
\pi_{B*} {\cal O}_\sigma(2 \sigma) \rightarrow 
R^1\pi_{B*} {\cal O}_B(\sigma) \rightarrow 
R^1\pi_{B*} {\cal O}_B(2 \sigma)
\ldots 
\eeq
By virtue of \eref{r1-1}, \eref{pi-2} and
\eref{induct1}, this sequence then reads
\beq
0 \rightarrow {\cal O}_{\IP^1} \rightarrow
\pi_{B*} {\cal O}_B(2 \sigma)  \rightarrow
{\cal O}_{\IP^1}(-2) \rightarrow 0 \ ,
\eeq
which implies that
\beq
\pi_{B*} {\cal O}_B(2 \sigma) = {\cal O}_{\IP^1} \oplus {\cal
O}_{\IP^1}(-2) \ .
\eeq
The pattern of induction is now clear. Continuing onwards we find,
in general, that
\beq
\label{pimsig}
\pi_{B*} {\cal O}_B(m \sigma) = {\cal O}_{\IP^1} \oplus {\cal
O}_{\IP^1}(-2) \oplus \ldots \oplus {\cal O}_{\IP^1}(-m)
\eeq
for any positive integer $m$.
Therefore, we see from \eref{piBCB} and \eref{pimsig} that
\beq
\pi_{B*} {\cal O}_B({\cal C}_B) = \left( {\cal O}_{\IP^1} \oplus
\bigoplus_{j=2}^m {\cal O}_{\IP^1}(-j) \right) \otimes {\cal
O}_{\IP^1}(k) \ .
\eeq
It then follows from expression \eref{h0push} that 
\begin{eqnarray}
\label{Cpiecegen}
h^0(B,{\cal O}_B({\cal C}_B)) 
&=& 
h^0(\IP^1, \left({\cal O}_{\IP^1} \oplus \bigoplus_{j=2}^m {\cal
	O}_{\IP^1}(-j) \right)\otimes {\cal O}_{\IP^1}(k)) \nn \\ 
&=&h^0(\IP^1, {\cal O}_{\IP^1}(k) \oplus \bigoplus_{j=2}^m 
	{\cal O}_{\IP^1}(-j+k)) \nonumber \\
&=& (k+1)+ \frac12 (2k - m)(-1 + m) \nonumber \\
&=& 1 - \frac12 m^2 + km + \frac12 m \ .
\end{eqnarray}
In the above calculation, we have used the fact that
\beq
h^0(\IP^1, {\cal O}_{\IP^1}(n)) = n+1 \mbox{~~~for~~~~}n \ge 0 \ . 
\eeq 
Note, from equation \eref{km}, that $k \ge m$. Therefore, 
expression \eref{Cpiecegen} is always non-negative, as it
must be.
Let us compare  \eref{Cpiecegen} with the special case where ${\cal
O}_B(\cC_B)$ is ample. 
This result is given in equation \eref{Cpiece} in
Appendix A. We see that they agree.
This is a very re-assuring consistency check.

\subsubsection{Spectral Line Bundle Moduli}
Next, we must compute $h^1({\cal C}_B, {\cal O}_{{\cal C}_B})$, the
number of
spectral line bundle moduli. This is none other than
the genus $g$ of the curve ${\cal C}_B$, that is,
\beq
h^1({\cal C}_B, {\cal O}_{{\cal C}_B}) = g 
\ .
\eeq 
To find the genus, we use the adjunction formula
\cite{hart1}
\beq
\label{adj}
2g - 2 = {\cal C}_B \cdot ({\cal C}_B + K_B) \ .
\eeq
We will henceforth use $K_B$ and $c_1(K_B)$ in adjunction formulas
interchangeably without ambiguity.
Recalling from \eref{dP9K} that $K_B = {\cal O}_B(-f)$,
we can use \eref{intersect} and \eref{coverbase} 
to compute \eref{adj} and obtain
\beq
2g -2 = -m^2+2km - m \ .
\eeq
Hence
\beq
\label{gcoverB}
h^1({\cal C}_B, {\cal O}_{{\cal C}_B}) = 1 - \frac12 m^2 +km - 
\frac12 m \ .
\eeq
Having computed both $h^0(B, {\cal O}_B({\cal C}_B))$ and $h^1({\cal
C}_B, {\cal O}_{{\cal C}_B})$, we can now determine dim$(IV)$.
Inserting \eref{Cpiecegen} and \eref{gcoverB} into
\eref{Bmoduli}, we find that
\beq
\label{termIVgen}
\dim(IV) = 1 - m^2 + 2km \ .
\eeq
Note that we actually know more than just the dimension of the moduli
space. Using the spectral data, we see that the structure of the moduli
space is actually
\beq
H^1(X,\pi^*M \otimes (\pi^*M)^*) = H^1(B,M \otimes M^*) \ .
\eeq
Equation \eref{Bmoduli} then
tells us that this moduli space is a direct product of a projective
space with a torus. 

The above results were computed directly within the
context of $B =  d\mathbb{P}_9$. However, if 
one wishes to know only the
dimension of the moduli space, then
one can give a more general computation
using the Atiyah-Singer index theorem that is applicable to any
base surface $B$. We present this computation 
in Appendix B. Happily, we see that the result obtained 
using the index theorem and given in \eref{h1AS} agrees with the
expression \eref{termIVgen}. Again, this is a very re-assuring
consistency check.


\subsection{Terms II and III: The Cross Terms}


We now turn to the computation of the two cross terms
$II = H^1(X, V \otimes (\pi^*M)^*)$ and 
$III= H^1(X, \pi^*M \otimes V^*)$.
We find that computing each term individually is very
difficult. However, by relating the two terms, we can arrive at
the requisite expressions for
$\dim(II)$ and $\dim(III)$. 
In this subsection, we will proceed as follow.
We first use the index theorem to
compute the difference $\dim(II) - \dim(III)$. Next, using Leray
spectral sequences, we will show that $\dim(II)$ and
$\dim(III)$ each counts the number of 
global holomorphic sections on some sheaf over a 
support curve in the base. We will calculate the degrees of these
sheafs and see that they can not both be simultaneously
positive. This means
that one of the two terms is always zero. Consequently, from our
expression for the
difference $\dim(II) - \dim(III)$, we can deduce expressions for 
$\dim(II)$ and $\dim(III)$, the terms we want.

%
\subsubsection{The Difference Between Terms II and III}
First, we find the difference between $\dim(II)$ and $\dim(III)$.
We use the Atiyah-Singer index theorem for $V \otimes (\pi^*M)^*$ in
term II, which implies that
\beq
\label{eulerC}
\chi(X, V \otimes (\pi^*M)^*)
= \int_X \ch(V \otimes (\pi^*M)^*) \wedge \td(TX)
= \sum_{i=0}^3 (-1)^i h^i(X, V \otimes (\pi^*M)^*) \ .
\eeq
Now, 
\beq
\label{h0-van}
h^0(X, V \otimes (\pi^*M)^*) = 0
\eeq
since $V$ is stable and, hence,
there are no global sections. It then follows that
\beq
\label{h3-van}
h^3(X, V \otimes (\pi^*M)^*) = 0
\eeq
by Serre duality. 
We can also use Serre duality to show that
\beq
H^2(X, V \otimes (\pi^*M)^*) = H^1(X, K_X\otimes(V \otimes
(\pi^*M)^*)^*) 
= H^1(X, \pi^*M \otimes V^*) = III \ .
\eeq
We have used the fact that $K_X$ is trivial since $X$ is a
Calabi-Yau manifold and the definition of term III given in
\eref{defII-III}. Therefore, using \eref{eulerC},
\eref{h0-van} and \eref{h3-van},
\beq
\label{eulerC2}
\int_X \ch(V \otimes (\pi^*M)^*) \wedge \td(TX)
= -\dim(II) + \dim(III) \ .
\eeq
Recall that for any vector bundle $U$,
\beq
\ch(U) = \mbox{rk}(U) + \ch_1(U) + \ch_2(U) + \ch_3(U) + \ldots ,
\eeq
where
\bea
\label{Chern}
\ch_1(U) &=& c_1(U), \quad
\ch_2(U) = \frac12(c_1(U)^2-2c_2(U)), \nn \\
\ch_3(U) &=& \frac16(c_1(U)^3 - 3 c_1(U)c_2(U) + 3 c_3(U)) \ , \ldots
\ .
\eea
Similarly,
\beq
\td(U) = 1 + \td_1(U) + \td_2(U) + \td_3(U) + \ldots \nn \\
\eeq
with
\beq
\label{Todd}
\td_1(U) = \frac12 c_1(U), \quad
\td_2(U) = \frac{1}{12} (c_2(U)+c_1(U)^2), \quad
\td_3(U) = \frac{1}{24} (c_1(U)c_2(U)) \ ,
\ldots \ .
\eeq
Using these expressions,
the left hand side of \eref{eulerC2} becomes
\beq
\int_X (n - c_2(V) + \frac12 c_3(V))\wedge (m - c_2(\pi^* M))\wedge
(1 + \frac{1}{12}c_2(TX)) \ ,
\eeq
which is equal to 
\beq
\frac12 m \int_X c_3(V) \ .
\eeq
This result and \eref{chernV} imply that expression
\eref{eulerC2} becomes
\beq
\label{diff23}
\dim(II) - \dim(III) = - m \lambda \eta \cdot (\eta- n f) \ .
\eeq

\subsubsection{Simplification for Term II}
Having obtained the difference between the expressions for
terms II and III in
\eref{diff23}, we proceed to simplify each term
individually. We will see that each counts the global holomorphic
sections of a sheaf on some support curve in the base. We will
then determine whether such sections exist by calculating the degree of
this sheaf.

Let us address term II first. As in the calculation of
term IV, we can fit term II into a Leray spectral sequence
\beq
0 \rightarrow H^1(B, \pi_* (V \otimes (\pi^* M)^*)) \rightarrow II
\rightarrow H^0(B, R^1\pi_*(V \otimes (\pi^* M)^*)) \rightarrow
H^2(B,\pi_* (V \otimes (\pi^* M)^*)) \rightarrow 
\ldots
\eeq
which, using the projection theorem 
\beq
R^i \pi_* (V \otimes (\pi^* M)^*) = R^i \pi_* V \otimes M^* \qquad
\mbox{ for }i \ge 0 \ ,
\eeq
becomes
\beq
\label{seqII-1}
0 \rightarrow H^1(B, \pi_* V \otimes M^*) \rightarrow II
\rightarrow H^0(B, R^1\pi_* V \otimes M^*) \rightarrow
H^2(B,\pi_* V \otimes M^*) \rightarrow 
\ldots
\eeq
We will first simplify \eref{seqII-1} substantially by arguing that
$\pi_* V$ vanishes. Let us rewrite $\pi_*V$ in a more useful
form. For some point $b$ in the base $B$,
we denote the sheaf of sections of $\pi_*V$ at $b$ by
$(\pi_*V)|_b$. Then, it is straight-forward to see that
\beq
(\pi_* V) |_b = V|_{\pi^{-1}(b)} = V|_F = H^0(F, V|_F) \ ,
\eeq
where $F$ is the fiber class on $X$, which we recall is an elliptic
curve,  
and $V|_F$ means the restriction of $V$ to
the fiber. We have used here the sheaf-theoretic interpretation of $V$
which allows us to conveniently write $(\pi_* V)|_b$ in terms of $H^0(F,
V|_F)$.
Now, the $n$-fold spectral cover of $V$ given in
\eref{coverX} intersects $F$ precisely 
$n$ times. We denote these points by $p_i$, with $i = 1, 2, \ldots,
n$. Moreover, let the zero section $\sigma$
intersect $F$ once at the point $e$. Over a generic point $b \in B$,
$e$ is distinct from the $n$ points $p_i$.
Therefore, at such a generic point, we can write, 
\beq
(\pi_* V)|_b = H^0(F, V|_F) = 
\bigoplus\limits_{i=1}^n H^0(F,{\cal O}_{F}(e - p_i)) \ . 
\label{pushV}
\eeq
Now, each bundle ${\cal O}_{F}(e - p_i)$ is clearly of degree zero,
being the sheaf for 
the divisor of points $e - p_i$. It is also holomorphic. 
However, one can show \cite{griffith} that a nontrivial
bundle of degree zero
over an elliptic curve admits no global sections.
Therefore,
$H^0(F,{\cal O}_{F}(e-p_i))$ is trivial for each $i$ and, hence,
it follows from \eref{pushV} that
\beq
(\pi_* V)|_b = 0 \ .
\eeq
But $V$ is torsion free, which implies that
$\pi_* V$ is also torision free \cite{hart1}. Therefore, $(\pi_* V)|_b
= 0$ for generic points $b \in B$ means that
\beq
\label{piv0}
\pi_* V = 0
\eeq
everywhere. Thus, our sequence \eref{seqII-1} reduces to
\beq
\label{seqII}
0 \rightarrow II \rightarrow H^0(B, R^1\pi_* V \otimes
M^*)
\rightarrow 0 \ ,
\eeq
which implies that
\beq
II \simeq H^0(B,R^1\pi_* V \otimes M^*) \ .
\eeq
Therefore,
\beq
\label{shortII}
\dim(II) = 
h^0(B,R^1\pi_* V \otimes M^*) \ .
\eeq

We would now like to evaluate $R^1\pi_*V$.
By definition III.8 of \cite{hart1}, we
have
\beq
R^1 \pi_* V = H^1(F, V|_F) \ .
\eeq
Furthermore, one can show that
\beq
\label{piv1}
\mbox{rk} R^1 \pi_* V = \mbox{rk} H^1(F, V|_F) = 0 \ .
\eeq
Using arguments similar to the above, we see that
\beq
(R^1\pi_* V)|_b = 0
\eeq
for the generic points $b \in B$ where $e$ is distinct from the $n$
points 
$p_i$. However, we can not conclude, as we did for $\pi_*V$, that
$R^1\pi_* V$ vanishes everywhere.
This is because the first
higher direct image functor, $R^1 \pi_*
V$, is not necessarily torsion free, even if $V$ is \cite{hart1}.
It follows that at any point $b' \in B$ over which
$e$ is equal to one of the points $p_i$,
$(R^1 \pi_* V)|_{b'}$ need not vanish. The locus of such special points
$b'$ form a co-dimension one object in $B$, namely, a curve.
This support curve, which we will denote by $C$, is given by
\beq
\label{CB}
C = \pi_*(\cC_X \cdot \sigma) \ .
\eeq
The sequence \eref{seqII} is clearly trivial everywhere except on
this curve. Therefore, any non-zero contribution to dim$(II)$ arises
from restricting the sheaf $R^1\pi_* V \otimes M^*$ to the curve $C$.
Note, using \eref{etaconv}, that $C$ is smooth.
Let us be more specific about the form of $C$.
Recalling the expression for $\cC_X$ from \eref{coverX} 
and using \eref{dP9K} and \eref{sigsig1}, the
curve $C$ defined in \eref{CB} becomes
\bea
\label{CB1}
C = \pi_*(\cC_X \cdot \sigma) &=& \pi_*( \pi^*(-n f + \eta) \cdot
	\sigma) \nn \\
	&=& \eta -n f \ .
\eea

When restricted to the support curve $C$,
\eref{shortII} 
becomes
\beq
\label{shortIIC}
\dim(II) = h^0(C, (R^1\pi_* V \otimes M^*)|_C) \ ,
\eeq
which is the number of the global holomorphic
sections of the sheaf $(R^1\pi_* V \otimes
M^*)|_C$ on $C$.
To determine the number of global sections, it will suffice to compute
the degree,
\beq
\label{deg}
d = c_1 \! \left((R^1\pi_* V \otimes M^*)|_C \right) \ ,
\eeq
of $(R^1\pi_* V \otimes M^*)|_C$. We will later use the fact that if
this degree is negative then $h^0(C, (R^1\pi_* V \otimes M^*)|_C)$
vanishes \cite{griffith}.

%
\subsubsection{The Degree $d$ Associated to Term II}
We now proceed to determine the degree $d$ in \eref{deg}. 
To do this, we invoke
the Groethendieck-Riemann-Roch theorem, which states
that for any map $f : X \rightarrow B$
and any sheaf $\cS$ on $X$, we have
%
\beq
\label{thmGRR}
\td (TB) \mbox{ch}(\sum_{i=0}^2 (-1)^i R^if_* \cS)
=
f_*(\mbox{ch}(\cS) \td(TX)) \ .
\eeq
For the case at hand, $\cS$ is the vector bundle $V$. Then,
\eref{thmGRR} becomes
\beq
\td(TB) \mbox{ch}(R^0\pi_* V - R^1\pi_* V)
= \pi_*(\mbox{ch}(V) \td(TX)) \ .
\eeq
From \eref{piv0}, we know that
$R^0\pi_* V = \pi_* V = 0$. Therefore, this expression simplifies to
\beq
\label{GRR1}
\td(TB) \mbox{ch}(-R^1\pi_* V) =
\pi_* (\mbox{ch}(V)\td(TX)) \ .
\eeq
Using \eref{Chern} and \eref{Todd}, we can expand \eref{GRR1} as
\bea
\label{GRRexp}
\left(1 + \td_1(TB) + \td_2(TB) \right)
\left(\ch_{1}(-R^1\pi_* V) +
\ch_{2}(-R^1\pi_* V) \right) = \qquad \nn \\
\pi_* \left(
\left( n + \ch_{2}(V) +\ch_{3}(V) \right)
\left( 1 + \td_2(TX) \right)
\right) \ .
\eea
In writing \eref{GRRexp},
we have used the facts that rk$(V) = n$,
$\td_1(TX) = \td_3(TX) = 0$ since $X$ is a Calabi-Yau manifold and
rk$(-R^1\pi_* V)=0$ by \eref{piv1}. Furthermore, terms on the left
hand side must terminate at order
2 since $B$ is of dimension 2 
and terminate 
at order 3 on the right hand side since $X$ is of dimension 3.
Multiplying out \eref{GRRexp}, we find that
\bea
\label{GRRexp1}
\ch_{1}(-R^1\pi_* V) + \ch_{2}(-R^1\pi_* V) + \ch_{1}(-R^1\pi_* V) 
\td_1(TB) \nn \\
=
\pi_*\left(
	n + n \td_2(TX) + \ch_{2}(V) + \ch_{3}(V)
\right) \ .
\eea
Using \eref{Chern} and \eref{Todd}, and identifying terms of equal
order, \eref{GRRexp1} implies that
\beq
\label{ch1ch2-1}
\ch_{1}(-R^1\pi_* V) = 
\pi_*\left(\ch_{2}(V) + n \ \td_2(TX) \right) \ , \quad
\ch_{2}(-R^1\pi_* V) = 
\pi_* \ch_{3}(V) - \td_1(TB)\ch_{1}(-R^1\pi_* V) \ .
\eeq
It then follows from \eref{chernX}, \eref{chernV}, the intersection 
\eref{intersect}, and using the fact that
$\pi_* F$ vanishes, that
\beq
\label{ch1ch2}
\ch_{1}(-R^1\pi_* V) = n f - \eta \ , \quad
\ch_{2}(-R^1\pi_* V) = \lambda \eta \cdot (\eta- n f) +
	\frac12 f \cdot \eta \ .
\eeq

Having obtained these results, recall that
we want to compute $c_1(R^1\pi_* V \otimes M^*)|_C)$. 
For convenience, let us define the sheaf
\beq
\label{defF}
\cF = R^1\pi_* V \otimes M^* \ .
\eeq
Next, we recall
the multiplicative property of the Chern character, 
that is, for sheafs $A$ and $B$
\beq
\label{chmulti}
\mbox{ch}(A \otimes B) = \mbox{ch}(A) \mbox{ch}(B) \ .
\eeq
Thus, from \eref{defF}, we have that
\beq
\label{cFchern}
\ch(-\cF) = \ch(-R^1\pi_* V)\ch(M^*) \ ,
\eeq
which, using \eref{Chern} and \eref{Todd} can be expanded into
\bea
\label{cFchern2}
&&\rk(-\cF) + \ch_{1}(-\cF) + \ch_{2}(-\cF) \nn \\
&&= 
\left(
  \rk(-R^1\pi_* V) + \ch_{1}(-R^1\pi_* V) + \ch_{2}(-R^1\pi_*V) 
\right)
\left(
  \rk(M^*) + \ch_{1}(M^*) + \ch_{2}(M^*) 
\right) \ . \nn \\
\eea
We can now make use of the facts that 
$\rk(-R^1\pi_* V) = 0$ from \eref{piv1},
$\rk(M^*) = \rk(M) = m$ and that $c_1(M) = 0$ from
\eref{chernM}. 
Identifying terms of order 1 and 2
respectively, \eref{cFchern2} becomes
\bea
\label{cFchern3}
\rk(-\cF) &=& 0 \nn \\
\ch_{1}(-\cF) &=& m \ \ch_{1}(-R^1\pi_* V) = m(n f - \eta) \\
\ch_{2}(-\cF) &=& m \ \ch_{2}(-R^1\pi_* V) 
= m(\lambda \eta \cdot (\eta- n f) +
	\frac12 f \cdot \eta) \nn \ ,
\eea
where we have used \eref{ch1ch2}.
We wish to compute $c_1(\cF|_C)$. To do this,
let us invoke the Groethendieck-Riemann-Roch theorem 
again, this time for the inclusion map
\beq
i : C \rightarrow B
\eeq
and the sheaf $-\cF = -R^1\pi_* V \otimes M^*$. Then
\beq
\label{GRRC}
i_* \left(
\td (TC) \ch (-\cF|_C)
\right)
=
\ch(\sum_{j=0}^2 (-1)^j R^j i_*(-\cF|_C) \td(TB) \ .
\eeq
However, all the higher image functors $R^j i_*$ for $j > 0$ vanish
because $i$ is an inclusion map. 
Moreover, $R^0 i_*(-\cF|_C) = i_*(-\cF|_C)$ is simply $-\cF$.
Therefore, \eref{GRRC} becomes
\beq
\label{GRRC1}
i_*\left( \td (TC) \ch(-\cF|_C) \right) 
=
\ch(-\cF) \td(TB) \ .
\eeq
Thus, expanding \eref{GRRC1} and using the first expresssion in
\eref{cFchern3}, we have
\bea
\label{GRRC1exp}
&&i_*\left((1 + \td_1(TC))(\ch_{0}(-\cF|_C)+\ch_{1}(-\cF|_C))
	\right) \nn \\
&=& 
\left(\ch_{1}(-\cF) + \ch_{2}(-\cF) \right)
\left(1 + \td_1(TB) + \td_2(TB) \right) \ .
\eea
Upon identifying terms of order 1 and 2 respectively, this implies
\beq
\label{rkchF-1}
\rk(-\cF|_C) i_*(1) = \ch_{1}(-\cF)
\eeq
and
\beq
\label{rkchF-2}
\ch_{1}(-\cF|_C) + \ch_{0}(-\cF|_C) \td_1(TC) =
	\ch_{1}(-\cF) \td_1(TB) + \ch_{2}(-\cF) \ .
\eeq
Noting that
\beq
i_*(1) = C \ ,
\eeq
and using the middle equation in \eref{cFchern3}, it follows from
\eref{rkchF-1} that
\beq
\rk(-\cF|_C) C = -m(\eta - n f) \ ,
\eeq
This then implies that
\beq
\label{rkandC}
\rk(-\cF|_C) = -m, \qquad C = \eta - n f \ .
\eeq
Both of these expressions are re-assuring. Note that 
\beq
\rk(-\cF|_C) = \rk(-R^1\pi_*V)\rk_C(M^*) = - m
\eeq
which is consistent with the first equation in \eref{rkandC}.
Second, the result for $C$ in \eref{rkandC} is identical to that found
in \eref{CB1}. Emboldened, let us proceed to
equation \eref{rkchF-2} 
which would give us what we are after. 
Now, we note that even though $\cF$ is not a vector bundle,
$\cF|_C$, being supported on $C$, is.
We recall that for any vector bundle $U$,
\beq
\label{invertc}
c_i(U^*) = (-1)^i c_i(U) \ .
\eeq
Therefore, from \eref{rkchF-2} and \eref{invertc}, 
we have
\beq
c_1(\cF|_C) = -c_1(-\cF|_C) = 
	-\left(\ch_{1}(-\cF) \td_1(TB) \frac{}{} + \ch_{2}(-\cF) 
	+ m \td_1(TC) \right) \ .
\eeq
Using \eref{cFchern3}, we find that
\bea
\label{ch1ch2-2}
c_1(\cF|_C) 
&=& - m \lambda \eta \cdot (\eta- n
	f) - m \ \td_1(TC) \ .
\eea
Also, we know that
\beq
\td_1(TC) = \frac12 c_1(TC) = \frac12 \chi(C) = \frac12 (2 -
2g(C)) \ ,
\eeq
where $g(C)$ is the genus of the curve $C$.
Recalling from \eref{deg} and \eref{defF} that
$d = c_1 \! \left(\cF|_C \right)$, we find 
from \eref{ch1ch2-2} that
\beq
\label{deg1}
d = - m \lambda \eta \cdot (\eta- n f) - m(1-g(C)) \ .
\eeq

We now need to find the genus of $C$.
Since $C$ is in $B$, to obtain the genus we again 
invoke the adjunction formula \eref{adj} in the base $B$,
\beq
g(C) = \frac12 C \cdot (C + K_{B}) + 1 \ .
\eeq
From \eref{etadot}
and \eref{CB1}, we have
\beq
C \cdot C = (\eta-n f)^2 = -2n \eta\cdot f + \eta^2 \ .
\label{CBCB}
\eeq
Similarly, we have
\beq
C \cdot K_{B} = (\eta-n f) \cdot (-f)
= - \eta \cdot f \ .
\label{CBK}
\eeq
Therefore, we find that
\beq
\label{gC}
g(C) = \frac12(\eta^2 -(2n+1) \eta\cdot f) + 1 \ .
\eeq

Upon substitution of \eref{gC} into \eref{deg1}, we find the degree
\beq
\label{deg2}
d = \frac{m}{2}((2\lambda-1)\eta^2 + (1+2n-2n\lambda) \eta \cdot f) \
.
\eeq
It is convenient to recast this expression into one
depending on the base curve $\eta'$ whose definition was given in
\eref{etaconv}. Doing this, \eref{deg2} becomes
\beq
\label{degII}
d = \frac{m}{2}\left(
(1-2\lambda) \eta'^2 + (1-2 (n+2) \lambda) \eta' \cdot f
\right) \ .
\eeq

%
%
\subsubsection{The Degree $d'$ Associated with Term III}
Having computed, in \eref{degII}
the degree $d$ of the sheaf $(R^1\pi_* V \otimes
M^*)|_C$ arising in term II, we will now compute
the analogous quantity for term III, which we will denote by $d'$. 
The
combined knowledge of $d$ and $d'$ will allow us to determine,
in conjunction with the index theorem result \eref{diff23},
the individual expression for $\dim(II)$ and $\dim(III)$.

First, we recall that $III = H^1(X, \pi^*M \otimes V^*)$.
We can fit this into a Leray
spectral sequence in precise analogy to what was done for term II.
Indeed, term III will be again supported on some curve because the
properties of $V$ used earlier, namely that $\pi_*V$ vanishes and
$R^1\pi_*V$ vanishes except on a curve, hold for the dual
bundle $V^*$ as well. Furthermore, the spectral cover for $V^*$ is
identical to that of $V$ \cite{Ron}. Thus, \eref{CB} implies 
that the support curve for $R^1\pi_* V^* \otimes M$ 
remains $C$ given in \eref{CB}.
Therefore, we have the analogue of equation \eref{shortIIC},
\beq
\dim(III) = h^0(C, (R^1\pi_* V^* \otimes M)|_C) \ .
\eeq
This is the number of global holomorphic sections of the 
sheaf $R^1\pi_* V^*\otimes M$ restricted to $C$. 
To find this, we need to determined the degree
\beq
\label{degIII1}
d' = c_1\left( (R^1\pi_* V^* \otimes M)|_C \right) \ .
\eeq

We repeat the analysis of the previous subsections, 
and find that the only
change in our expression \eref{deg} is that we now have
$\ch_{3}(V^*)$ rather than $\ch_{3}(V)$. Using
\beq
\ch_{3}(V^*) = -\ch_{3}(V) \ ,
\eeq
\eref{degIII1} is found to be
\beq
\label{degIII1-2}
d' =  m \lambda \eta \cdot (\eta- n f) - m(1-g(C)) 
\ .
\eeq
In terms of the curve $\eta'$ defined in \eref{etaconv},
\eref{degIII1-2} becomes
\beq
\label{degIII}
d' = \frac{m}{2}\left(
(1+2\lambda) \eta'^2 + (1+2 (n+2) \lambda) \eta' \cdot f
\right) \ .
\eeq

%
\subsubsection{Comparing $d$ and $d'$}
We now compare \eref{degII} and \eref{degIII}.
First, we note that $\eta'^2$ and $\eta' \cdot f$ are both positive by
our constraints \eref{etaample1} and \eref{etaample2}. 
Therefore, if $\lambda \ge \frac12$,
then $d < 0$ while $d' > 0$. Now, because there are no global
holomorphic section to a sheaf of negative degree, this would imply
that $\dim(II)$ vanishes. On the other hand, if $\lambda \le -\frac12$,
then $d > 0$ while $d' < 0$. This would imply that  $\dim(III)$
vanishes.
This leaves the one single case of $\lambda=0$, 
for which our arguments
do not apply.
When $\lambda=0$, both $d$ and $d'$ are positive,
meaning that both $\dim(II)$ and $\dim(III)$ are non-zero. However,
their difference in this case, by \eref{diff23}, is zero. Therefore,
in this special case $\dim(II) = \dim(III)$.
For convenience, let us henceforth assume that
\beq
\label{lambdacons}
\lambda \ne 0 \ .
\eeq
In summary, then
\bea
\label{d-2-3}
\lambda \ge \frac12, && \dim(II) = 0 \nn \\
\lambda \le -\frac12, && \dim(III) = 0 \ .
\eea
That is, one of the two cross terms always vanishes
if $\lambda \ne 0$.
%
%
\subsubsection{Obtaining $\dim(II)$ and $\dim(III)$}
Combining our discussions in the previous subsections, we
at last can obtain explicit expressions for
$\dim(II)$ and $\dim(III)$.
Applying \eref{d-2-3} with the index theorem result \eref{diff23}, we
find that
\beq
\label{diff23-1}
\dim(II) = 0 \ , \quad
\dim(III) = m \lambda \eta \cdot (\eta - n f), 
\eeq
for $\lambda \ge \frac12$ and
\beq
\label{diff23-2}
\dim(II) = -m \lambda \eta \cdot (\eta - n f) \ , \quad
\dim(III) = 0
\eeq
for $\lambda \le -\frac12$.
It is re-assuring to see that all cases are non-negative, as they must
be. Using \eref{etadot}, the individual expressions
\eref{diff23-1} and \eref{diff23-2}
can be combined to give
\beq
\label{termII-III}
\dim(II) + \dim(III) = m |\lambda|
\left(
b^2 - 3 n b + \sum_{i=1}^9 (a_i - a_i^2)
\right)
\eeq
for $\lambda \ne 0$.

\section{Final Result: The Moduli for $\hat{V} = V \oplus \pi^*M$}
Collecting all four terms from \eref{termI}, \eref{termIVgen},
and \eref{termII-III}, 
we have our final result for the dimension of the moduli space of a
reducible rank $n+m$ holomorphic
vector bundle of the form $\hat{V} = V \oplus \pi^* M$ on a Calabi-Yau
threefold $X$ elliptically fibered over $d{\mathbb P}_9$. 
If
we denote the moduli space of this bundle by ${\cal M}( V \oplus \pi^*
M )$, then 
\bea
\label{moduli}
\dim \left({\cal M}(V\oplus \pi^*M)\right) &=& 
(m |\lambda| + \frac12 n) b^2 + 
(3 - \frac32 n - 3mn|\lambda|) b  + 
(n + 2km - m^2) + \nn \\
&&\sum_{i=1}^9 \left((1+m|\lambda|-\frac12 n^2 )a_i - 
	(\frac12 n + m |\lambda|) a_i^2 \right) \ ,
\eea
where 
\beq
\ba{rcll}
\lambda &=& p + \frac12, \qquad &n \mbox{~~odd} \\
\lambda &=& p, \qquad & n \mbox{~~even}
\ea
\eeq
with $p \in \IZ$,
\beq
\lambda \ne 0
\eeq
and
\beq
1 < m \le k  \ .
\eeq
The parameters $a_i$ and $b$ are 
defined by
\beq
a_i = a_i' - (n+1), \quad b = b' + 3(n+1) \ ,
\eeq
where $a_i'$ and $b'$ obey the following constraints
\beq
\label{pos2}
b'^2-\sum_{i=1}^9 a_i'^2 > 0, \qquad
3b'+ \sum\limits_{i=1}^9 a_i' > 0, \qquad
b'\beta - \sum\limits_{i=1}^9 a_i' \alpha_i > 0
\eeq
for all $\alpha_i \in \IZ$ and $\beta \in \IZ_{>0}$ such that
\beq
- \beta^2 + \sum\limits_{i=1}^9 \alpha_i^2 = 1, \qquad 
3 \beta + \sum\limits_{i=1}^9 \alpha_i = 1 \ .
\eeq

To illustrate these results, let us present a sample calculation.
An obvious solution to the positivity constraints \eref{pos2}
is
\bea
a_i' &=& -n, \qquad i = 1,2,\ldots,9 \nn \\
b' &=& 3n + 1 \ . 
\eea
This implies that
\bea
a_i &=& -2n-1, \qquad i = 1,2,\ldots,9 \nn \\
b&=&6n+4 \ .
\eea
Substituting these coefficients into
result \eref{moduli} gives
\beq
\dim({\cal M}(V\oplus \pi^*M)) =
3 + 2\,k\,m - 2\,|\lambda|\,m - m^2 
- 18\,|\lambda|\,m\,n - 18\,|\lambda|\,m\,n^2 
+ \frac{9\,n^2}{2} + \frac{9\,n}{2}  \ .
\eeq


%
%
\section*{Acknowledgements}
We are grateful to R.~Donagi and T.~Pantev for many insightful
comments.
This 
Research was supported in part under
cooperative research agreement \#DE-FG02-95ER40893
with the U.~S.~Department of Energy and an NSF Focused Research Grant
DMS0139799 for ``The Geometry of Superstrings''.

%
\section*{Appendix A: Ample Case for the Spectral Curve Moduli of Term
IV}
In this Appendix, we compute $h^0(B, {\cal O}_B({\cal C}_B) )$ for the
simple case where ${\cal O}_B({\cal C}_B)$ is ample. We will compare
the result with the general case \eref{Cpiecegen} computed in the text
of this paper and find that they agree.

Consider the bundle ${\cal O}_B({\cal C}_B) \otimes K_B^{-1}$ and
assume that it is ample.
Then its positivity allows us to invoke the Kodaira vanishing
theorem \cite{griffith}
\beq
H^q(B, \Omega^p({\cal O}_B({\cal C}_B) \otimes K_B^{-1})) = 0,
\qquad \mbox{for } p+q > \dim B \ .
\eeq
Taking $p=2$ and using the fact that $\Omega^p({\cal O}_B({\cal
C}_X)) = \Omega^2(TB) \otimes {\cal O}_B({\cal C}_B) = K_B \otimes {\cal
O}_B({\cal C}_B)$, we have, for all $q > 0$, that
\beq
H^q(B, {\cal O}_B({\cal C}_B)) =
H^q(B,\Omega^2(TB) \otimes {\cal O}_B({\cal C}_B) \otimes K_B^{-1})
= 0 \ .
\eeq
Therefore
\beq
\chi(B, {\cal O}_B({\cal C}_B) ) 
= \sum_{i=0}^2 (-1)^i h^i(B, {\cal O}_B({\cal C}_B) ) 
= h^0(B, {\cal O}_B({\cal C}_B) ) \ .
\eeq
Now,
\beq
\chi(B, {\cal O}_B({\cal C}_B) ) =
\int_B \mbox{ch}({\cal C}_B) \wedge \td(TB)
\eeq
by the Groethendieck-Riemann-Roch theorem. 
For surfaces $B$, this reduces to
\bea
h^0(B, {\cal O}_B({\cal C}_B) ) &=&
\frac12 {\cal C}_B \cdot ( {\cal C}_B - K_B) + \frac1{12} (K_B^2 +
c_2(TB) )
\nonumber \\
&=& \frac12 {\cal C}_B \cdot {\cal C}_B - \frac12 {\cal C}_B \cdot K_B +
\frac1{12} c_2(TB) \ ,
\eea
where we have used the fact, from \eref{cherndP9},
that $K_B^2 = 0$ on a $d{\mathbb P}_9$.
We also recall from \eref{c2dP9} that $c_2(TB) = 12$,
from \eref{fiberdP9} that ${\cal C}_B = m\sigma_B + k f$ and from
\eref{dP9K} that $c_1(K_B) = -f$. Together with
the intersection numbers given in \eref{intersect}, we find that
\beq
\label{Cpiece}
h^0(B, {\cal O}_B({\cal C}_B) ) = 1 - \frac12 m^2 + km + \frac12 m \ .
\eeq
This expression indeed agrees with the result for the general case of
${\cal O}_B({\cal C}_B)$ computed in \eref{Cpiecegen}.


%
\section*{Appendix B: $\dim(IV)$ from the Index Theorem}
In this Appendix we calculate the dimension of term IV, that is,
$h^1(B, M \otimes M^*)$, using the Atiyah-Singer index theorem. 
We then compare the result to
\eref{termIVgen} computed in the $d{\mathbb P}_9$ context
and find perfect agreement.

First, we note that $H^0(B, M \otimes M^*) = 1$ because $M$ is
stable. 
Then, by Serre duality we have
\beq
H^2(B, M \otimes M^*) = H^0(B, K_B \otimes M^* \otimes M) \ ,
\eeq
which vanishes by the arguments following \eref{thirdIV}.
Therefore,
the Atiyah-Singer index theorem
\beq
\chi(B, M \otimes M^*) = \sum_{i=0}^2(-1)^i
h^i(B, M \otimes M^*)
= \int_B \mbox{ch}(M \otimes M^*) \wedge \td(TB) \ ,
\eeq
now reduces to
\beq
h^1(B, M \otimes M^*) =
1 - \int_B \mbox{ch}(M \otimes M^*) \wedge \td(TB) \ .
\label{h1chiB}
\eeq
Using \eref{chernM}, \eref{Chern} and \eref{chmulti}, we have
\beq
\label{chMM}
\ch(M \otimes M^*) = (m - k \ \mbox{pt})\wedge(m - k \ \mbox{pt})
= m^2 - 2 k m \ \mbox{pt} \ ,
\eeq
where we have been careful in including the class pt of points.
Similarly, from \eref{cherndP9}, \eref{c2dP9} and \eref{Todd}, we have
\beq
\label{toddTB}
\td(TB) = 1 + \frac12 f + 1 \ \mbox{pt} \ .
\eeq
Multiplying \eref{chMM} with \eref{toddTB},
and inserting into \eref{h1chiB}, we obtain the final result
\beq
\label{h1AS}
\dim(IV) = 
h^1(B, M \otimes M^*) 
= 1 - m^2 + 2 k m \ ,
\eeq
which is in perfect agreement with \eref{termIVgen}.

\bibliographystyle{JHEP}

\end{document}